\documentclass[Physsubmission, Phys]{SciPost}
\hypersetup{
    colorlinks,
    linkcolor={red!50!black},
    citecolor={blue!50!black},
    urlcolor={blue!80!black}
}

\newcommand{\non}{\nonumber\\}
\usepackage{amsmath,amsthm,amssymb,amsfonts,empheq}
\usepackage[bitstream-charter]{mathdesign}
\usepackage{enumitem}

\DeclareSymbolFont{usualmathcal}{OMS}{cmsy}{m}{n}
\DeclareSymbolFontAlphabet{\mathcal}{usualmathcal}

\begin{document}

% TODO: write your article's title here.
% The article title is centered, Large boldface, and should fit in two lines
\begin{center}{\Large \textbf{
QCD factorization for hadronic quarkonium production at high $p_T$\\
}}\end{center}

% TODO: write the author list here. Use initials + surname format.
% Separate subsequent authors by a comma, omit comma at the end of the list.
% Mark the corresponding author with a superscript *.
\begin{center}
Kyle Lee\textsuperscript{1,2},
Jian-Wei Qiu\textsuperscript{3,4}, 
George Sterman\textsuperscript{5}, and
Kazuhiro Watanabe\textsuperscript{3,6$\star$}
\end{center}

% TODO: write all affiliations here.
% Format: institute, city, country
\begin{center}
{\bf 1} Nuclear Science Division, Lawrence Berkeley National Laboratory, Berkeley, CA 94720, USA
\\
{\bf 2} Physics Department, University of California, Berkeley, CA 94720, USA
\\
{\bf 3} Theory Center, Jefferson Lab, Newport News, Virginia 23606, USA
\\
{\bf 4} Department of Physics, The College of William \& Mary, Williamsburg, Virginia 23187, USA
\\
{\bf 5} C.N.~Yang Institute for Theoretical Physics and Department of Physics and Astronomy, 
Stony Brook University, Stony Brook, NY 11794, USA
\\
{\bf 6} SUBATECH UMR 6457 (IMT Atlantique, Universit\'e de Nantes, IN2P3/CNRS), 4 rue Alfred Kastler, 44307 Nantes, France
\\
% TODO: provide email address of corresponding author
* watanabe@jlab.org
\end{center}

\begin{center}
\today
\end{center}

% For convenience during refereeing (optional),
% you can turn on line numbers by uncommenting the next line:
%\linenumbers
% You should run LaTeX twice in order for the line numbers to appear.

\definecolor{palegray}{gray}{0.95}
\begin{center}
\colorbox{palegray}{
  \begin{tabular}{rr}
  \begin{minipage}{0.1\textwidth}
    \includegraphics[width=22mm]{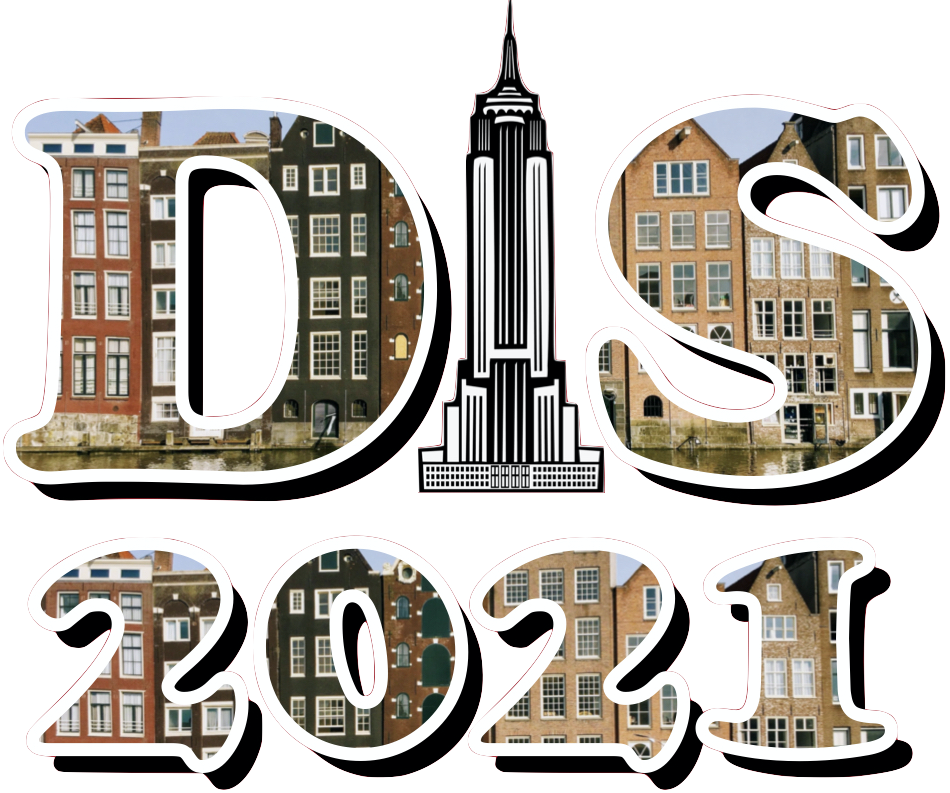}
  \end{minipage}
  &
  \begin{minipage}{0.75\textwidth}
    \begin{center}
    {\it Proceedings for the XXVIII International Workshop\\ on Deep-Inelastic Scattering and
Related Subjects,}\\
    {\it Stony Brook University, New York, USA, 12-16 April 2021} \\
    \doi{10.21468/SciPostPhysProc.?}\\
    \end{center}
  \end{minipage}
\end{tabular}
}
\end{center}

\section*{Abstract}
{\bf
% TODO: write your abstract here.
%The abstract is in boldface, and should fit in 8 lines.
%It should be written in a clear and accessible style, emphasizing the context, the problem(s) studied, the methods used, the results obtained, 
%the conclusions reached, and the outlook. You can add a table contents, recommended if your paper is more than 6 pages long.

Heavy quarkonium production at high transverse momentum ($p_T$) in hadronic collisions is explored in the QCD factorization approach.  We find that the leading power in the $1/p_T$ expansion is responsible for high $p_T$ regime, while the next-to-leading power contribution is necessary for the low $p_T$ region. 
We present the first numerical analysis of the scale evolution of coupled twist-2 and twist-4 fragmentation functions (FFs) for heavy quarkonium production and demonstrate that the QCD factorization approach is capable of describing the $p_T$ spectrum of hadronic $J/\psi$ production at the LHC. 

}

% TODO: include a table of contents (optional)
% Guideline: if your paper is longer that 6 pages, include a TOC
% To remove the TOC, simply cut the following block
%\vspace{10pt}
%\noindent\rule{\textwidth}{1pt}
%\tableofcontents\thispagestyle{fancy}
%\noindent\rule{\textwidth}{1pt}
%\vspace{10pt}

%%%%%%%%%%%%%%%%%%%%%%%%%%%%%%%%%%%%%%%%%%%%%%%%%%%%%%%
\section{Introduction}
\label{sec:intro}
% TODO: write your article here.
%The stage is yours. Write your article here.
%The bulk of the paper should be clearly divided into sections with short descriptive titles, including an introduction and a conclusion.

Understanding heavy quarkonium production is a challenging and exciting research subject in the study of QCD. NRQCD factorization~\cite{Bodwin:1994jh} has successfully described many features of existing data. However, at the LHC energies, significant enhanced contributions in powers of $\ln(p_T^2/m^2)$ are not fully included in fixed order NRQCD calculations, affecting the shape of calculated $p_T$ spectrum of heavy quarkonium production. 

The renormalization group improved QCD factorization approach is capable of studying such logarithmically enhanced higher-order contributions systematically~\cite{Nayak:2005rt,Kang:2014tta}. 
The QCD factorization approach expands the $p_T$ spectrum in powers of $1/p_T$ first. It factorizes both the leading power (LP) and next-to-leading power (NLP) contributions in terms of perturbatively calculable hard parts (expanded in powers of $\alpha_s$) convoluted with universal parton distribution functions (PDFs) and fragmentation functions (FFs).  This approach is reliable if the uncertainty from all neglected contributions in higher powers of $\alpha_s$ and higher powers of $1/p_T$ are controllable, and its predictive power relies on the universality of the PDFs and FFs.  This paper shows that the QCD factorization approach can describe the prompt $J/\psi$ production in hadronic collisions ($A+B\to J/\psi(p)+X$) at the LHC energies, and argue that both LP and NLP contributions are essential for describing the shape of $J/\psi$'s $p_T$ spectrum.

\begin{figure}[t]
\centering
\includegraphics[width=0.495\textwidth]{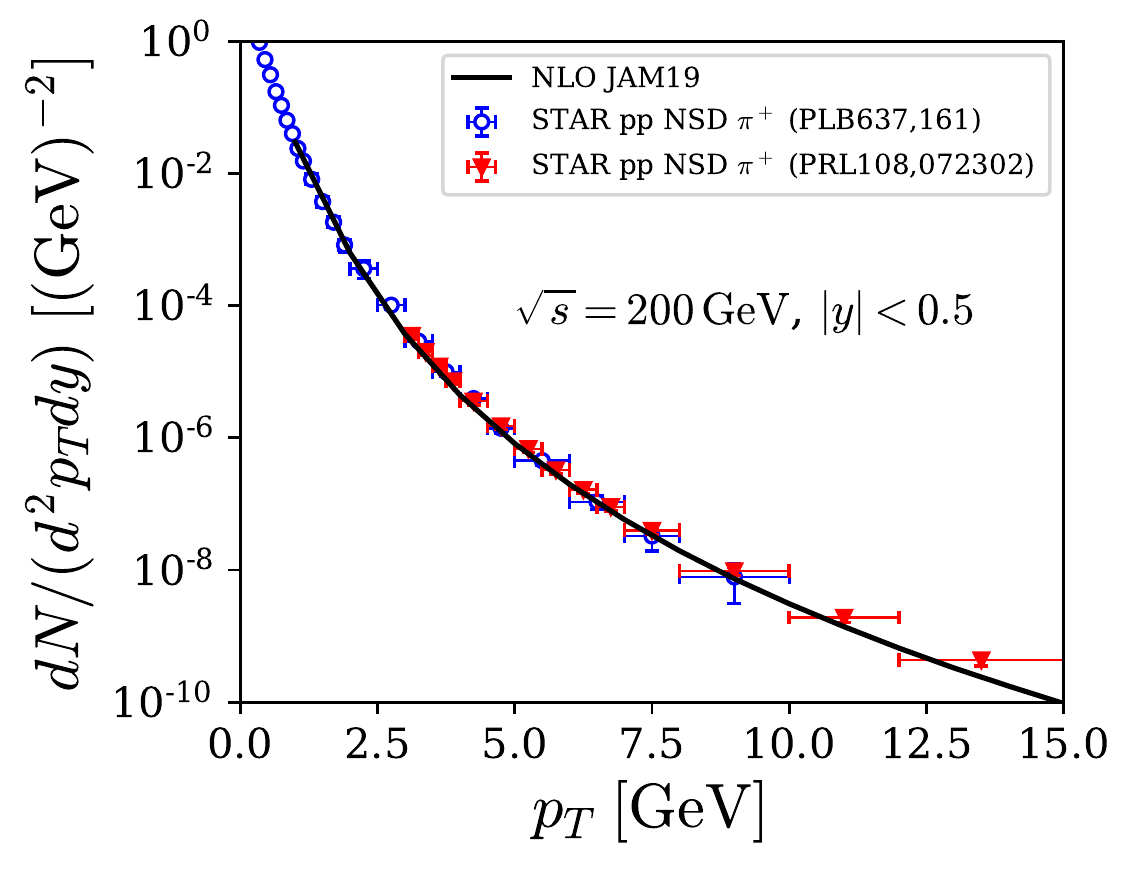}
\includegraphics[width=0.495\textwidth]{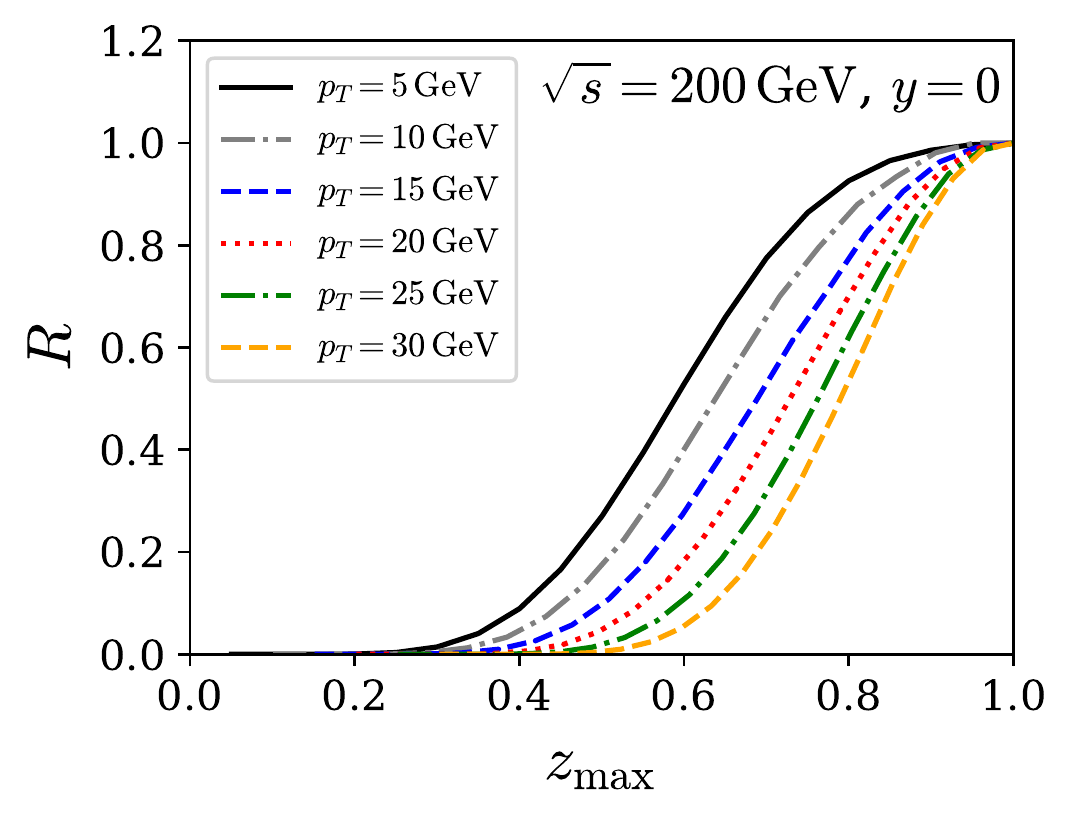}
\caption{(Left): Differential cross section for $\pi^+$ production in $p+p$ collisions at mid rapidity at RHIC.
Data are from Refs.~\cite{STAR:2006xud,STAR:2011iap}. 
(Right): The normalized 
hadronic $\pi^+$ production as a function of $z_{\rm max}$, see text for the definition of $R$ and $z_{\rm max}$.}
\label{fig:pip-RHIC}
\end{figure}

%%%%%%%%%%%%%%%%%%%%%%%%%%%%%%%%%%%%%%%%%%%%%%%%%%%%%%%
\section{Quarkonium production in QCD factorization}
\label{sec:qcd-factorization}

%=========================================================================
\subsection{LP contribution}
\label{subsec:LP}

Inclusive production of a single hadron of mass $m_h$ at high $p_T$ ($\gg m_h$) in hadronic collision can be factorized in QCD as~\cite{Nayak:2005rt},
\begin{align}
E_p\frac{d\sigma_{A+B\to H(p)+X}}{d^3p}\bigg|_{\rm LP}
= \sum_{f=u,d,s,c}\int_{z_{\rm min}}^1 \frac{dz}{z^2}D_{f\to H}(z,\mu^2)
E_c\frac{d\hat{\sigma}_{A+B\to f(p_c)+X}}{d^3p_c}\left(p_c=\frac{p}{z},\mu^2\right),
\label{eq:LP}
\end{align}
where $\hat{\sigma}_{A+B\to f(p_c)+X}$ represents the cross section to produce the fragmenting parton of flavor $f$ and momentum $p_c$ with all collinear sensitivities around $p_c\sim p/z$  absorbed into the twist-2 parton-to-hadron FFs, $D_{f\to H}$ with momentum fraction $z$ and factorization scale $\mu$, and can be further factorized into PDFs of colliding hadrons and perturbatively calculable hard parts, which are available at the next-to-leading (NLO) accuracy in $\alpha_s$ expansion~\cite{Aversa:1988vb}. Corrections to \eqref{eq:LP} are expected to be suppressed by the power of $1/p_T$.

The factorization formula in \eqref{eq:LP} has been successful in interpreting data on light hadron production, such as STAR-data for $\pi^+$ production in $p+p$ collisions at RHIC, as shown in Fig.\,\ref{fig:pip-RHIC} (Left). 
The theory curve was obtained by using JAM19 sets~\cite{Sato:2019yez} for PDFs and $\pi^+$ FFs with 
$\mu^2=p_T^2$.
We find a nice agreement between the theoretical curve and data points for $p_T\gtrsim 1\,{\rm GeV}$, which indicates that $\ln(p_T^2/m^2)$-type logarithmically enhanced contributions start to dominate when $p_T/m \gtrsim 5$ (or 7) with $m\sim \Lambda_{\rm QCD}$ (or $m\sim m_{\pi}$) and power corrections in $m/p_T$ are sufficiently small.  We note that
high $p_T$ hadron production in $p+p$ collisions is more sensitive to the FFs at large $z$ (in comparison with the production in $e^+ e^-$ or $e^- p$ collisions) due to the steep falling nature of PDFs of two colliding hadrons at large momentum fraction $x$. To quantify this feature, we plot 
$R\equiv \left[{\int_{z_{\rm min}}^{z_{\rm max}}dz/z^2 D_{f\to \pi^+}d\hat{\sigma}}\right]
/\left[{\int_{z_{\rm min}}^{1}dz/z^2 D_{f\to \pi^+}d\hat{\sigma}}\right]$ 
in Fig.\,\ref{fig:pip-RHIC} (Right), where about 50\% of the cross section results from $z=0.7$ and above at $p_T=15\,{\rm GeV}$, while $D_{f\to \pi^+}$ is falling fast when $z$ increases.  This feature is specially relevant to heavy quarkonium production since its fragmentation functions are likely peaked in the large $z$ region.

The factorization formalism in \eqref{eq:LP} 
should be applicable to $J/\psi$ production when $D_{f\to \pi}$ is replaced by $D_{f\to J/\psi}$, so long as the power corrections in $1/p_T$ are sufficiently small and the $\ln(p_T^2/m^2)$-type contributions dominate the production cross section~\cite{Nayak:2005rt}. Since it is necessary to have a $c\bar{c}$ pair to form a $J/\psi$, the fragmenting parton should have a minimum virtuality, $m\gtrsim 2m_c \gg m_\pi$. If we require the similar dominance of $\ln(p_T^2/m^2)$-type contributions to the $\pi$ production, we expect the formula in \eqref{eq:LP} to work for $J/\psi$ production when $p_T \gtrsim 5\, (\text{or}\, 7)\, 2m_c \sim 15-20$\,GeV. Since producing a high $p_T$ $c\bar{c}$ pair at the hard collision is suppressed by $1/p_T^2$ comparing the production of single fragmenting parton, the LP formalism in \eqref{eq:LP} covers only events where $c\bar{c}$ pairs emerge at distances longer than  $1/\mu_0$ with $\mu_0 \sim 2m_c$ - the scale of non-perturbative input FFs, $D_{f\to J/\psi}(z,\mu_0^2)$. As shown in Sec.\,\ref{sec:method}, the factorized LP contribution provides a good description of the published LHC data at high $p_T\gtrsim 60$\,GeV, but is far below the data when extrapolated to lower $p_T$.

%=========================================================================
\subsection{NLP contribution}
\label{subsec:NLP}

The NLP contributions to the inclusive production of a single hadron at high $p_T$ can also be factorized and could be particularly important for heavy quarkonium production~\cite{Kang:2014tta}: 
\begin{align}
E_P\frac{d\sigma_{A+B\to H(p)+X}}{d^3p}\bigg|_{\rm NLP}\approx 
\sum_{\kappa}\int \frac{dz}{z^2} D_{[Q\bar{Q}(\kappa)]\to H}(z,\mu^2)
E_c\frac{d\hat{\sigma}_{A+B\to [Q\bar{Q}(\kappa)](p_c)+X}}{d^3p_c}\left(p_c=\frac{p}{z},\mu^2\right),
\label{eq:NLP}
\end{align}
where $\hat{\sigma}_{A+B\to [Q\bar{Q}(\kappa)](p_c)+X}$ represents the cross section to produce a fragmenting $Q\bar{Q}$ pair of spin-color state $\kappa$ and momentum $p_c = P_Q+P_{\bar{Q}} = P'_Q+P'_{\bar{Q}}$, where $P'_Q$ and $P'_{\bar{Q}}$ are momenta in the conjugated production amplitude, with all collinear sensitivities around $p_c$ absorbed into the twist-4 $Q\bar{Q}(\kappa)$-to-hadron FFs, $D_{[Q\bar{Q}(\kappa)]\to H}$.  For simplicity, in this paper, we approximate $P_Q=P_{\bar{Q}}=P'_Q=P'_{\bar{Q}}=p/(2z)$~\cite{Kang:2014tta}. Although corresponding partonic hard parts to produce a pair of heavy quarks are $1/p_T^2$ suppressed, the NLP contribution could be important since it is more likely to get the quarkonium from a fragmenting $Q\bar{Q}$-pair than a single fragmenting parton~\cite{Kang:2014tta}.  With the $1/p_T^2$ suppressed hard parts at LO, derived in Ref.~\cite{Kang:2014pya}, as shown in Sec.\,\ref{sec:method}, we find that the NLP contribution provides the much needed enhancement at low $p_T$ to improve the overall description of the LHC data from the QCD factorization approach.

%=========================================================================
\subsection{Renormalization group improvement}
\label{subsec:evolution}

Physically observed cross sections should not depend on the factorization approach to describe them.  Renormalization group improved QCD factorization at the NLP accuracy requires the twist-2 and twist-4 FFs to satisfy the following coupled evolution equations~\cite{Kang:2014tta},
\begin{align}
\frac{\partial}{\partial \ln\mu^2} D_{f\to H}(z,\mu^2)
=&\, \frac{\alpha_s(\mu)}{2\pi}\sum_{f'}\int_z^1\frac{dz'}{z'}P_{f\to f'}\left(\frac{z}{z'}\right)D_{ f' \to H}(z',\mu^2)
\non
+&\,\frac{\alpha_s^2(\mu)}{\mu^2}\sum_{\kappa}\int_z^1\frac{dz'}{z'} P_{f\to[Q\bar{Q}(\kappa)]}\left(\frac{z}{z'}\right)
D_{[Q\bar{Q}(\kappa)]\to H}\left(z',\mu^2\right)\,,
\label{eq:twist2-evolution}
\end{align}
%\\
\begin{align}
\frac{\partial}{\partial \ln\mu^2} D_{[Q\bar{Q}(\kappa)]\to H}(z,\mu^2)
=&\frac{\alpha_s(\mu)}{2\pi}\sum_{n}\int^1_{z}\frac{dz'}{z'} P_{[Q\bar{Q}(n)]\to [Q\bar{Q}(\kappa)]}\left(\frac{z}{z'}\right)\, 
D_{[Q\bar{Q}(n)]\to H}(z',\mu^2)\,,
\label{eq:twist4-evolution}
\end{align}
where the first line of \eqref{eq:twist2-evolution} is the well-known DGLAP evolution of the twist-2 FFs as a consequence of requiring the renormalization group improved QCD factorization at the LP accuracy, and the second line of \eqref{eq:twist2-evolution} represents a NLP contribution to the DGLAP evolution, which effectively resums logarithmically enhanced contributions to the cross section when the produced fragmenting parton fragments to a heavy quark pair at a scale between $[\mu_0,\mu\sim p_T]$, and the pair then fragments to the observed quarkonium $H$.  In \eqref{eq:twist2-evolution}, the evolution kernels, $\alpha_s^2(\mu)P_{f\to[Q\bar{Q}(\kappa)]}\left(z\right)\equiv \gamma_{f\to[Q\bar{Q}(\kappa)]}\left(u=\frac{1}{2},v=\frac{1}{2},z\right)$ with $\gamma_{f\to[Q\bar{Q}(\kappa)]}$ given in Ref.~\cite{Kang:2014tta}.  In \eqref{eq:twist4-evolution}, 
the evolution kernels, \\
$\frac{\alpha_s(\mu)}{2\pi} P_{[Q\bar{Q}(n)]\to [Q\bar{Q}(\kappa)]}(z)\equiv \int_0^1du\int_0^1dv\,\Gamma_{[Q\bar{Q}(n)]\to [Q\bar{Q}(\kappa)]}\left(u,v,u'=\frac{1}{2},v'=\frac{1}{2}, z\right)$ with $\Gamma_{[Q\bar{Q}(n)]\to [Q\bar{Q}(\kappa)]}$ given in Ref.~\cite{Kang:2014tta}. 

At the NLP accuracy, the renormalization group improved and factorized cross section covers all events in which the heavy quark pair can be produced at the short-distance \eqref{eq:NLP}, at the input scale \eqref{eq:LP}, or in-between \eqref{eq:twist2-evolution}.  Unlike the power corrections to the cross section in \eqref{eq:NLP}, which go away by powers of $1/p_T^2$, the contribution to the cross section from the power correction to the evolution of twist-2 FFs in \eqref{eq:twist2-evolution} remains important even at large $p_T$ because its contribution to the cross section is built up from $\mu_0$ to $\mu \sim p_T$ and heavy quarkonium FFs are peaked in the large $z$ region.

\begin{figure}[t]
\centering
\includegraphics[width=\textwidth]{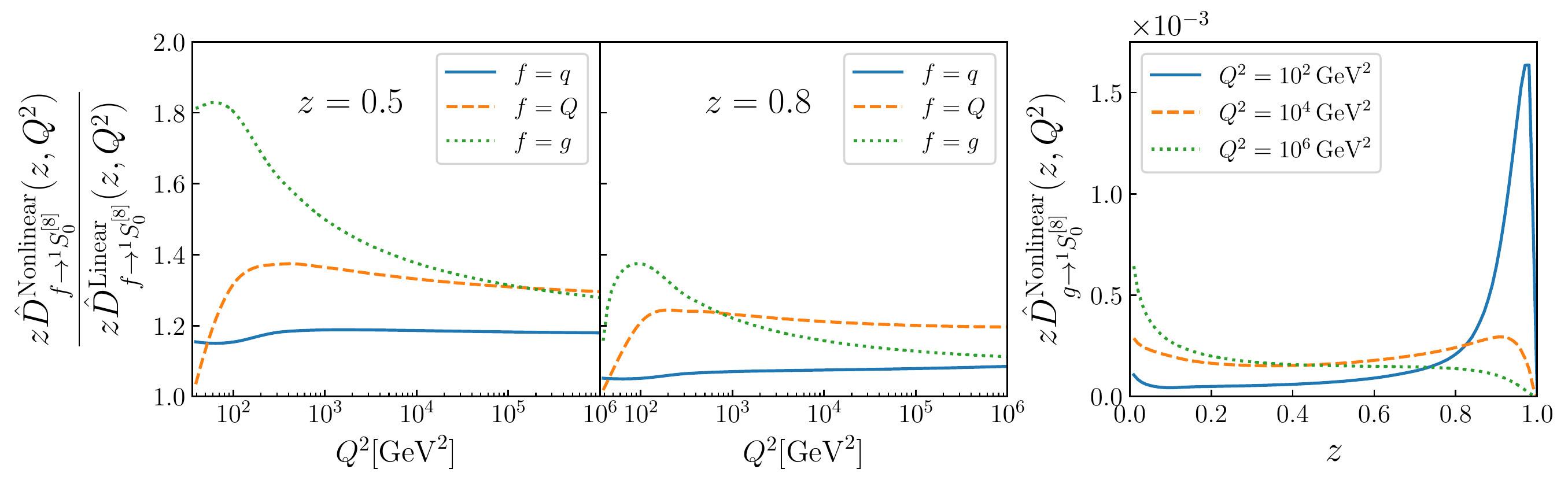}
\caption{(Left): Ratios of the LP FFs with the nonlinear corrections to that without the nonlinear corrections for $\hat{D}_{f\to {^1S_0^{[8]}}}\equiv D_{f\to H}/\langle \mathcal{O}_{^1S_0^{[8]}}^{H}\rangle$. 
(Right): The $\hat{D}_{g\to {^1S_0^{[8]}}}$
LP FF with the nonlinear corrections as a function of $z$. }
\label{fig:LP-FFs}
\end{figure}

%%%%%%%%%%%%%%%%%%%%%%%%%%%%%%%%%%%%%%%%%%%%%%%%%%%%%%%
\section{Numerical results for $J/\psi$ production}
\label{sec:method}

The predictive power of the QCD factorization approach to $J/\psi$ production at high $p_T$, combining Eqs.\,\eqref{eq:LP}-\eqref{eq:twist4-evolution}, relies on our knowledge of the non-perturbative twist-2 and twist-4 FFs.  With the heavy quark mass 
$m_c\gg \Lambda_{\rm QCD}$, as a model, we could apply NRQCD factorization to express the analytic twist-2 and twist-4 FFs at the input scale, $\mu_0\gtrsim 2m_c$ in terms of a small set of NRQCD long-distance-matrix-elements (LDMEs) with their $z$-dependence calculated perturbatively in NRQCD in an expansion of $\alpha_s$ and heavy quark velocity $v$ in the pair's rest frame~\cite{Nayak:2005rt,Kang:2014tta,Kang:2014pya}.  Both twist-2 and twist-4 FFs at the input scale $\mu_0$ have been derived for both LO and NLO, and expressed in terms of four LDMEs corresponding to the pair in spin-color states: ${^3S_1^{[1]}}$, ${^1S_0^{[8]}}$, ${^3S_1^{[8]}}$, ${^3P_J^{[8]}}$ with $J=0,\,1,\,2$~\cite{Ma:2013yla,Ma:2014eja}.  
In principle, one could solve the evolution equations in \eqref{eq:twist2-evolution} and \eqref{eq:twist4-evolution} with the NRQCD calculated input FFs at $\mu_0$, and use calculated hard parts and the QCD factorization formalisms in \eqref{eq:LP} and \eqref{eq:NLP} to predict the $J/\psi$'s $p_T$ spectrum at the LHC energies.

In practice, perturbatively calculated FFs are only well-defined under the integration over $z$ due to their dependence on (i) $\delta(1-z)$, (ii) $f(z)\ln(1-z)$ and (iii) $f(z)/[1-z]_+$ and $f(z)[\ln(1-z)/(1-z)]_+$ with $f(z)$ a regular function and the standard ``+'' prescription for [...]$_+$. As functions of $z$, these types of contributions to input FFs as perturbative coefficients of $\alpha_s^n$ with $n=1,2$ could be much larger than one, for example, as $z\to 1$, making the perturbative expansion not reliable.  Furthermore, gluon radiation to neutralize a fragmenting $c\bar{c}$ pair's color necessarily requires $D_{[Q\bar{Q}(\kappa)]\to J/\psi}(z) \to 0$ as $z\to 1$, while the $\delta(1-z)$ and $f(z)\ln(1-z)$ dependence from the fixed order perturbative calculations leads to an unphysical infinity as $z\to 1$.  Even under the integration, if we solve the evolution equations in terms of Mellin moments, the fact that the FFs dominate at the large $z$ requires special care for taking the inverse to get the evolved distributions as functions of $z$. 
Therefore, instead of worrying about the perturbative stability and size of higher order corrections, in this paper, we model these three types of contributions to the input FFs as 
$N\, z^\alpha (1-z)^\beta/B[1+\alpha,1+\beta]$, 
where $\alpha$ and $\beta$ are free parameters, $B$ is the Euler Beta-function, and $N$ is equal to the first moment of the corresponding term, which takes into account the relative size of different terms from perturbative calculations~\cite{Ma:2013yla,Ma:2014eja}.
If the first moments are negative, we take the absolute values to keep the same order of magnitude for the contributions.
For all other contributions that vanish at $z=1$, we use corresponding analytical expressions.  
With different choices of $(\alpha,\beta)$, we could make the $z$-dependent contribution peaked at any value of $z$.

\begin{figure}[t]
\centering
\includegraphics[width=0.495\textwidth]{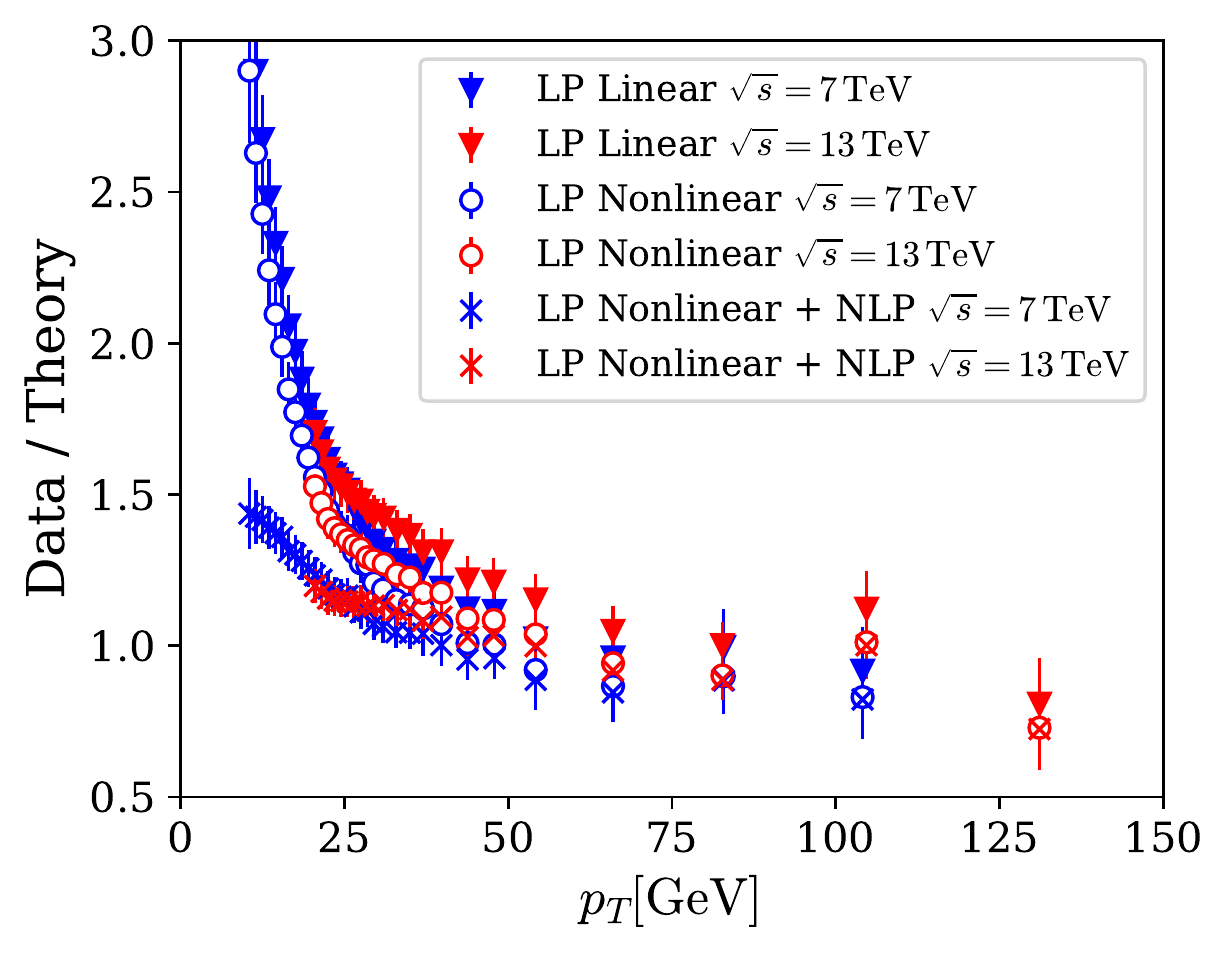}
\includegraphics[width=0.495\textwidth]{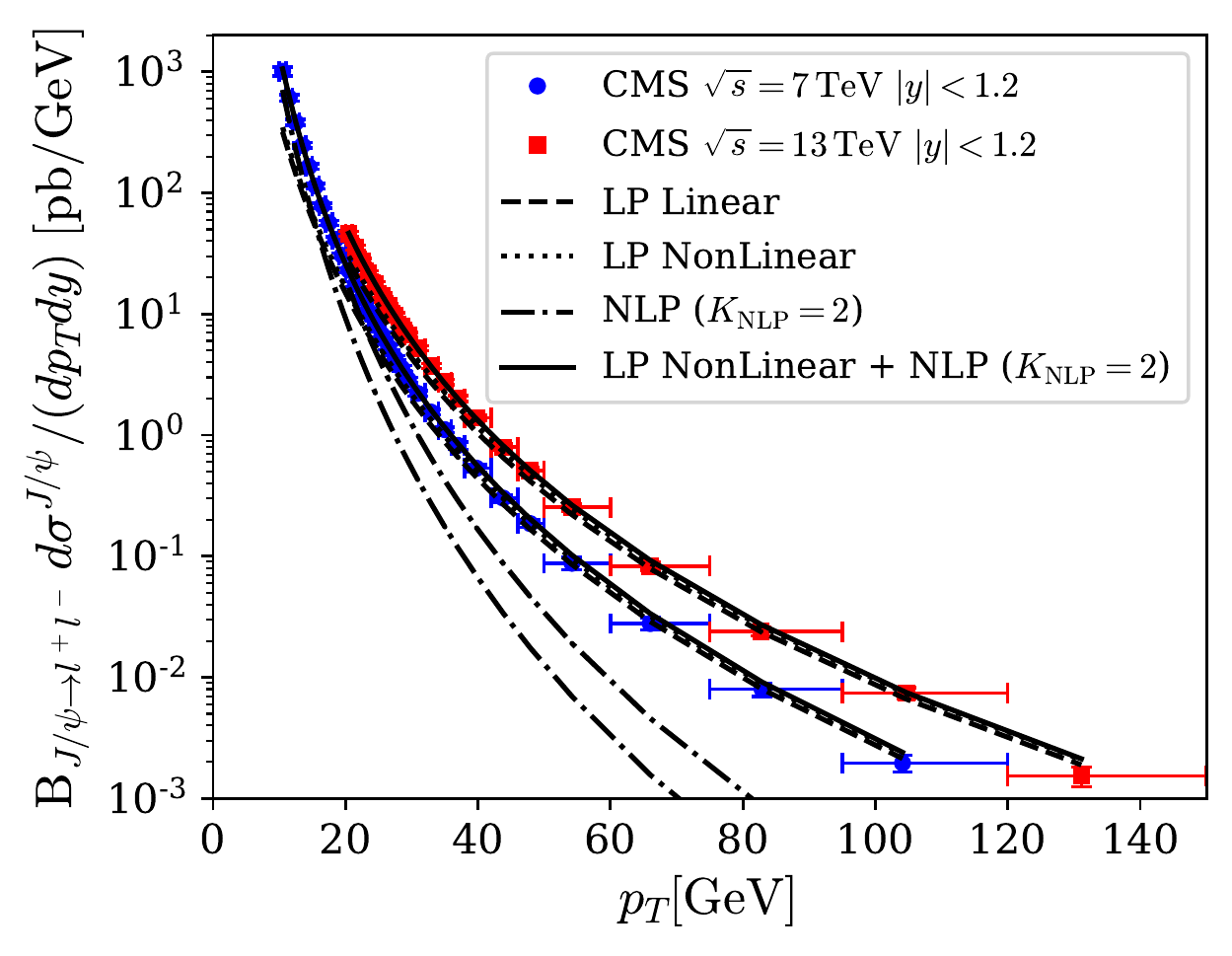}
\caption{(Left): Ratios of CMS data to theoretical calculations for hadronic $J/\psi$ production at the LHC. 
(Right): 
Prompt $J/\psi$ production in $p+p$ collisions at mid rapidity at the LHC with $K_{\rm NLP}=2$ and ($\alpha,\beta$) given in the text.
}
\label{fig:xsection}
\end{figure}

In our numerical calculations, we set $\alpha=30$, $\beta=0.5$ to have the input FFs peaked near $z=1$ for both the twist-2 and twist-4 input FFs at $\mu_0=4m$ with $m=m_{J/\psi}/2\approx 1.5\,{\rm GeV}$ a charm quark mass embedded in the input FFs. 
In Fig.\,\ref{fig:LP-FFs} (Left), we demonstrate that the impact of the nonlinear correction in \eqref{eq:twist2-evolution} does not disappear even at higher scales since the correction is to the evolution slope of FFs, not the FFs themselves~\cite{Mueller:1985wy}.  
As a result, the twist-2 FFs at large $z$ can be enhanced by about 10--30\% due to the nonlinear evolution even at a large probing scale.
In Fig.\,\ref{fig:xsection}, we compare our calculation with CMS data on prompt $J/\psi$ production in the rapidity bin $|y|<1.2$~\cite{Khachatryan:2015rra,Sirunyan:2017qdw}. 
We find that the production is dominated by the ${^1S_0^{[8]}}$ channel, and we fix the LDME $\langle \mathcal{O}_{^1S_0^{[8]}}^{J/\psi}\rangle$ by fitting the LP contribution without the nonlinear corrections to CMS data at $p_T=60\,{\rm GeV}$ and above at $\sqrt{s}=7,\,13\,{\rm TeV}$. Using CT18NLO set for PDFs~\cite{Hou:2019efy}, we obtained $\langle \mathcal{O}_{^1S_0^{[8]}}^{J/\psi}\rangle =0.129 \pm 5.18\times 10^{-3}\,{\rm GeV}^3$, which is similar to the one obtained by NLO fixed order NRQCD calculations~\cite{Chao:2012iv}. 
We show the ratios between CMS data and three sorts of theoretical results in Fig.\,\ref{fig:xsection}(Left).
The NLO LP contribution with the nonlinear corrections describes CMS data at high $p_T$, while the NLP contribution becomes more significant around $p_T=30\,{\rm GeV}$ and below.  Since we used the NLP partonic cross section at LO in our calculations, we could include a $K$-factor to mimic NLO contributions. In Fig.\,\ref{fig:xsection}(Right), a nice agreement between theoretical results and CMS data can be achieved with the chosen ($\alpha,\beta$) and $K_{\rm NLP}=2$.

%%%%%%%%%%%%%%%%%%%%%%%%%%%%%%%%%%%%%%%%%%%%%%%%%%%%%%%
\section{Conclusion}
\label{sec:conclusion}

We presented the first numerical calculations for $J/\psi$ production in hadronic collisions in the renormalization group improved QCD factorization formalism, including the NLP contribution. We demonstrated that the LP contributions dominate the high $p_T$ region, while the NLP contributions are sizable at lower $p_T$ and necessary for describing the LHC data within the QCD factorization approach.  With only two parameters ($\alpha,\beta$) and $K_{\rm NLP}$, theoretical calculations in terms of QCD factorization are consistent with existing data while there is sufficient room to improve.

\section*{Acknowledgements}
We wish to thank Hee Sok Chung and Nobuo Sato for valuable discussions and their instructions to use \texttt{INCNLO} codes regarding Ref.~\cite{Aversa:1988vb}. J.W.Q. and K.W. thank Nobuo Sato for teaching them a way to implement \texttt{LHAPDF6} interpolator~\cite{Buckley:2014ana}. 
K.W. would also like to thank Jefferson Lab for computational resources essential to perform this project. This work is supported by Jefferson Science Associates, LLC under U.S. DOE Contract No.\,DE-AC05-06OR23177.

% TODO: include author contributions
% TODO: include funding information
%\paragraph{Funding information}
%Authors are required to provide funding information, including relevant agencies and grant numbers with linked author's initials. Correctly-provided data will be linked to funders listed in the \href{https://www.crossref.org/services/funder-registry/}{\sf Fundref registry}.

% TODO:
% Provide your bibliography here. You have two options:

% FIRST OPTION - write your entries here directly, following the example below, including Author(s), Title, Journal Ref. with year in parentheses at the end, followed by the DOI number.
%\begin{thebibliography}{99}
%\bibitem{1931_Bethe_ZP_71} H. A. Bethe, {\it Zur Theorie der Metalle. i. Eigenwerte und Eigenfunktionen der linearen Atomkette}, Zeit. f{\"u}r Phys. {\bf 71}, 205 (1931), \doi{10.1007\%2FBF01341708}.
%\bibitem{arXiv:1108.2700} P. Ginsparg, {\it It was twenty years ago today... }, \url{http://arxiv.org/abs/1108.2700}.
%\end{thebibliography}

% SECOND OPTION:
% Use your bibtex library
% \bibliographystyle{SciPost_bibstyle} % Include this style file here only if you are not using our template

\bibliography{BibTexData_DIS21.bib}

\nolinenumbers

\end{document}